
\input epsf
%
\input harvmac

%
%
\ifx\epsfbox\UnDeFiNeD\message{(NO epsf.tex, FIGURES WILL BE IGNORED)}
\def\figin#1{\vskip2in}
\else\message{(FIGURES WILL BE INCLUDED)}\def\figin#1{#1}\fi
\def\ifig#1#2#3{\xdef#1{fig.~\the\figno}
\goodbreak\midinsert\figin{\centerline{#3}}%
\smallskip\centerline{\vbox{\baselineskip12pt
\advance\hsize by -1truein\noindent\footnotefont{\bf Fig.~\the\figno:}
#2}}
\bigskip\endinsert\global\advance\figno by1}

\def\ifigure#1#2#3#4{
\midinsert
\vbox to #4truein{\ifx\figflag\figI
\vfil\centerline{\epsfysize=#4truein\epsfbox{#3}}\fi}
\narrower\narrower\noindent{\footnotefont
{\bf #1:}  #2\par}
\endinsert
}


\Title{RU-94-93}{\vbox{\centerline{Modular Cosmology} }}
\smallskip
\centerline{\it
T. Banks  \footnote{*}
{\rm  John S. Guggenheim Foundation Fellow, 1994-95;
Varon Visiting Professor at the Weizmann Institute of Science;
Supported in part by the Department of Energy under grant No.
DE-FG05-90ER40559  .},M. Berkooz, S. H. Shenker \footnote{***}{\rm Supported in
part by the
Department of Energy under grant No. DE-FG05-90ER40559 .} }
\smallskip
\centerline{Department of Physics and Astronomy}
\centerline{Rutgers University}
\centerline{Piscataway, NJ 08855-0849}
\smallskip
\centerline{\it G. Moore \footnote{****}{\rm Supported in part by the
Department of Energy under grant DE-FG02-92ER40704, and by a
Presidential Young Investigator Award from the National Science
Foundation}}
\smallskip
\centerline{Department of Physics}
\centerline{Yale University}
\centerline{New Haven, CT 06520}
\smallskip
\centerline{\it P.J.Steinhardt\footnote{**}{\rm John S. Guggenheim
Fellow 1994-95, Supported in part by the Department of Energy under
grant No. DOE-EY-76-C-02-3071, by National Science Foundation Grant
NSF PHY 92-45317,
 , by Dyson Visiting Professor Funds of the Institute
for Advanced Study,
Princeton, and by NSF Grant PHY89-04035 at Santa
Barbara}\footnote{$\dagger$}{\rm On Leave at the Institute for
Theoretical Physics, Santa Barbara}}
\smallskip
\centerline{Department of Physics}
\centerline{University of Pennsylvania}
\centerline{Philadelphia, PA 19104}
\noindent
\bigskip
\baselineskip 18pt

\vfill
\noindent
We begin a study of the cosmology of moduli in string theory.  The
quantum field theory requirement of \lq\lq naturality'' is shown to be
incompatible with slow roll inflationary cosmology unless very stringent
constraints are satisfied.  In most cases, these constraints imply the
existence of fields with the properties of string moduli: they must have
only
nonrenormalizable couplings to light fields and their natural range of
variation must be the Planck scale.  The scale which characterizes their
potential energy (the inflation scale) must be two to three orders of
magnitude smaller than the Planck mass in order to explain the
observed magnitude of the fluctuations in the cosmic microwave
background.  Even if these constraints are
satisfied, generic initial conditions near the Plank energy density do
not lead to inflation unless the theory contains topological defects.
In this case inflation can arise naturally at the cores of the defects.
We show that string theory has two generic types of domain walls which
could be the seeds for inflation, and argue that modular physics
provides a very robust model of inflation.
Two scenarios are presented to explain the discrepancy between the
inflation scale and the scale of supersymmetry (SUSY) breaking.
One of them is favored because it leads to a natural understanding of
why the dilaton does not run out to the weak coupling region in the
postinflationary period
\Date{February 1995}
\eject

\newsec{Introduction}

The existing versions of superstring theory all contain high dimension
 manifolds of degenerate ground states.  As a consequence,
string theory contains massless fields which move on these ground state
 manifolds as functions of space and time.  For supersymmetric
ground states, the degeneracy is not lifted in perturbation theory and these
 massless {\it moduli} fields remain
massless to all orders in perturbation theory.  It is quite clear that nature
 does not contain such massless scalar fields,
and so one must hope that nonperturbative physics lifts the moduli masses up
 to an acceptable energy scale.
It is an implicit assumption of all work on string theory that this occurs.
  Generally it is also assumed that the
dynamics responsible for the masses of the moduli is connected with the
 dynamical breakdown of space time supersymmetry which
must occur if string theory is to describe the real world.

Among the many moduli one, the dilaton, is of special interest because it
 controls the string coupling.  General
 arguments\ref\dineseiberg{M.Dine, N.Seiberg, {\it Phys. Lett.}{\bf
 162B},(1985),299.}
show that it cannot be stabilized in a regime where a systematic weak coupling
 perturbation expansion of all of string theory is valid.
At least some sectors of the theory must be strongly coupled.  Most string
 theorists have kept this problem tucked away
in the back of their minds, assuming that the dilaton could be stabilized, and
 that somehow one would then explain why the
observed couplings in the world are weak at short distance.  There are
currently
two scenarios which purport to explain how this could occur. The first is the
``racetrack'' of \ref\race{L.Dixon, Talk Given at the 15th APS
Division of Particles and Fields Meeting, Houston, TX, Jan. 3-6, 1990
({\it This talk summarizes work of Dixon,Kaplunovsky and Peskin}),
SLAC-PUB 5229; N.V.Krasnikov, {\it Yad. Fiz.}{\bf 45},(1987),293.}.
It has proven difficult to find a SUSY
breaking vacuum state with zero cosmological constant within this framework.
The other proposal for stabilizing the dilaton\ref\banksdine{T.Banks,
M.Dine,
{\it Phys. Rev.}{\bf D50},(1994),7454.} invokes
 nonperturbative
stringy corrections\ref\shenker{S.Shenker, {\it The Strength of
Nonperturbative Effects in String Theory}, in the Cargese 90
Proceedings,
{\it Random Surfaces and Quantum Gravity}, Cargese, France, May 28 -
June 1 1990.} to the dilaton's Kahler potential, combined
with a superpotential generated by nonperturbative field theoretic effects.
Since the corrections to the Kahler potential are
presently uncalculable, it is hard to assess the plausibility of this proposal.

In the present paper we will find that a robust inflationary cosmology
can be constructed if we make some modest general assumptions about the
potential on moduli space.  If we were forced to live within the
straightjacket of race track models however, it is easy to show that
our ``modest general assumptions'' would be untenable.  For our purposes
then the proposal of \banksdine {}is at least a necessary psychological
crutch.  While we will not use any of the explicit results of that paper,
we {\it will} rely on the freedom it allows us in imagining the form of
the modular potential.

Recently, it has begun to become evident that
the scenarios for generation of moduli masses
tacitly assumed by most string theorists lead to problems
with cosmology.
In particular, Brustein and one of the present
authors\ref\problems{R.Brustein, P.J.Steinhardt, {\it Phys. Lett.}{\bf
B302},(1993),196.}
argued that even if the dilaton were stabilized at a finite value of the
 coupling, generic cosmological initial conditions
would send it flying out to the weak coupling region in which string theory
 conflicts with experiment.
In addition, moduli with masses set by the scale of SUSY breaking would, even
 in an inflationary universe, dominate the energy density
of the universe as nonrelativistic matter until it was too low for
 nucleosynthesis to take place\ref\bkn{T.Banks, D.Kaplan, A.Nelson, {\it
 Phys. Rev.}{\bf D49},(1994),779.}\ref\casas{B.DeCarlos, J.A.Casas,
 F.Quevedo, E.Roulet, {\it Phys. Lett.}{\bf B318},(1993),447.}.

Motivated by these difficulties, we have begun a study of the cosmology of
 moduli\foot{The proposal that moduli are the inflaton fields of inflationary
 cosmology dates back to work of Binetruy and
 Gaillard\ref\maryk{P.Binetruy, M.K.Gaillard, {\it Phys. Rev.}{\bf
 D34},(1986),3069. }.},
or what we will call
 {\it modular cosmology}.
It is our hope not only to solve these problems, but to find a version of
 cosmology in which moduli play an important role.
This might lead to an early opportunity to confront string theoretic
 predictions with observational data.
Although we have not yet succeeded in finding a satisfactory modular cosmology
we have obtained some interesting results.

In particular, if, following \ref\frieman{K.Freese, J.Frieman,
A.V.Olinto, {\it Phys. Rev. Lett.}{\bf 65},(1990),3233; F.C.Adams,
\break J.R.Bond, K.Freese, J.Frieman, A.V. Olinto, {\it Phys. Rev.}{\bf
D47},(1993),426.}, one incorporates the field
 theoretic
requirement of naturality on the lagrangians of inflationary cosmology, then
one is led to conclude that successful inflation requires the existence of
fields with the properties of moduli.  The argument is simple.  The number of
e-foldings of the universe in a slow roll
 inflationary model is given by\foot{Here we
 write
formulae for models involving a single field, but our
 conclusions are completely general.}
\eqn\ne{N_e \sim {1\over M_P^2}\int d\phi {V(\phi )\over V'(\phi )} }
We will argue that in natural models this number can be large only if
there exist fields whose range of variation is the Planck scale.

\lq\lq Naturalness'' or \lq\lq Technical Naturalness'' is a strong
constraint on scalar field lagrangians which follows from a combination
of general
renormalization group analysis and common sense.  It is basically the
requirement that all small parameters in a theory be explained by
well understood physical mechanisms.  For example, a parameter is
allowed to be small if it violates a symmetry whose breaking can be
attributed to effects of a very weak coupling or nonrenormalizable
terms in the lagrangian.

There are two classes of natural scalar field lagrangians.  The first class is
appropriate for a field with renormalizable interactions like the Higgs fields
of the standard model. Such lagrangians have the generic form
\eqn\renormlag{{\cal L} = \ha (\nabla\phi )^2 + \ha m^2 \phi^2 +
{\mu\over 3!}\phi^3 + {\lambda\over 4!}\phi^4 + o({1\over M^p})O_{4 +
p} }
  Here $M$ is a mass scale, much larger than $m$ or $\mu$ and
$O_d$ is an operator of dimension $d$.
The quadratic and cubic terms in the lagrangian of such a field
are unnaturally small, but this can be explained by SUSY or by technicolor-like
ideas.  The quartic couplings of such fields are of order one, and higher order
couplings are suppressed by powers of a very high energy scale.  Lagrangians
of this form never lead to inflation\foot{Actually, this statement is a bit
too strong.  Certain renormalizable supersymmetric models with very large
discrete symmetry groups may lead to successful inflationary
models\ref\micha{M.Berkooz, {\it Unpublished}
}.  Even in this class of models, inflation occurs for field values of
order the Planck scale and one must understand the high energy dynamics of
the theory.}.

The second
class of natural lagrangians has the form
\eqn\natlag{{\cal L} = [G_{ij}({\phi\over f})\partial_{\mu} \phi^i
g^{\mu\nu}\partial_{\nu} \phi^j - M^4 v({\phi \over f})]\sqrt{-g}}
This would be appropriate for a pseudo Goldstone boson of an accidental
symmetry
with decay constant $f$.  If the symmetry were broken only by Planck
scale effects, $M$ would be of order $f ({f\over M_P})^q$, where $q$
is a positive integer determined by the lowest dimension operator
which breaks the continuous symmetry.  The other example
of this sort of lagrangian is provided by the moduli fields of string theory.
In this case $f\sim M_P$ and $M \ll M_P$.

For this second class of lagrangian, $N_e$ is of order $({f\over M_P})^2
\int dx {v(x)\over v'(x)}$ .
  Thus, unless $f\sim M_P$ it is difficult to get any inflation
at all.  The only loophole in this argument is the possibility of a divergence
in the dimensionless integral.  If the divergence occurs at a finite point,
or at infinity in a noncompact space of finite volume, then $f$ must still
be of order $M_P$ to get sufficient inflation.  In this case one would argue
that the small probability of starting the universe off with initial conditions
near the point where the integral diverges\foot{One must remember that quantum
 mechanics
limits the degree to which we can assume very accurate classical initial
conditions.} is compensated by the large volume of the late time universe
covered by those regions with initial conditions which led to inflation.
This sort of {\it a postiori} calculation
 has been carried out by \frieman and others
.  It leads to a significant probability for inflation only when $f\sim M_P$.

There is one po
ssible exception to this argument\foot{T.B. thanks L.
Randall for a discussion of this point.}. If field
space is noncompact and of infinite volume, and if the potential approaches
a constant or increases like a small power (the precise power depends on $M$
and $f$), then one can have inflation at large field strengths for
Lagrangians of the form \natlag , even when $f\ll M_P$.  This is essentially
the scenario used by Linde\ref\lindechaos{A.Linde, {\it Phys. Lett.}{\bf
129B},(1983),177.}
 in his chaotic inflation models.  The small dimensionless constants necessary
to the
success of this scenario might be naturally realized as powers of a ratio
of scales.
Apart from this
possibility\foot{Note that this scenario is not appropriate for
string moduli, for the space of moduli fields has finite
volume \ref\moorehorne{J.Horne, G.Moore, {\it Nucl. Phys.}{\bf
B432},(1994),109.}. It is also inappropriate for Goldstone bosons, since
in that case field space is compact.  We do not really have examples in
which fundamental physical principles lead to fields with the properties
required for large field inflation.},
one is led to the conclusion that for field theoretically
natural lagrangians, inflation requires the existence of fields with a
Planck scale range of variation.  This observation is very exciting for
a string theorist, since it suggests that string theory moduli are the natural
candidates for
inflaton fields, a proposal first made by Binetruy and Gaillard \maryk {}.

There are two more simple observations that are relevant to our study of
modular cosmology.  The first is that even for $f\sim M_P$, one cannot hope
to get many e-foldings of inflation from a natural lagrangian.  Even the
$60$ or so e-foldings required to solve the standard cosmological puzzles,
will have to be explained by dimensionless numerical factors in the
equations.  For example, the dimensionless coefficient in the
expansion of the potential around some quadratic maximum might have to
be $ \sim  .01$ in order to achieve sufficient inflation.

The second is that the fluctuations in the microwave background
observed by the COBE experiment\ref\cobe{G.F. Smoot  {\it et al},
{\it Ap. J. Lett.}{\bf 396}, L1
(1992); C. L. Bennett {\it et al}, submitted to {\it Ap. J. Lett.} (1994);
 K. M. Gorski, {\it et al.},   {\it Ap. J.} {\bf 430},
L89 (1994); J.R. Bond, submitted to {\it Phys. Rev. Lett.} (1994).},
constrain the parameter $M$
to be about $10^{16} - 10^{17}$ $GeV$ (assuming that we wish to retain the
standard inflationary explanation of these fluctuations).  This follows
from
the formula
\eqn\fluct{10^{-5} \sim {\delta\rho\over \rho}\sim {H^2 \over
\dot{\phi}}\sim ({M\over M_P})^2}
Here we have used a standard formula for inflationary fluctuations
\ref\kolb{E.W.Kolb, M.S.Turner, {\it The Early Universe}, Addison-Wesley
Publishing Company, 1990, p. 287.}, and the slow roll equations of
motion, dropping various numerical coefficients.  The latter could
change our estimate of the fluctuations by an order of magnitude.
On the other hand, if SUSY is to solve the hierarchy problem, the natural scale
for the vacuum energy near the true minimum of the potential, the square of
the SUSY breaking $F$ term, is bounded by $10^{10} - 10^{11}$ $GeV$.  For
larger values
 of $F$, squark masses, which are generated by dimension six operators
neutral under all symmetries, will be larger than $1$ $TeV$, even if SUSY
breaking is communicated only by gravitational strength interactions.
We will present several speculative
explanations of the discrepancy between the inflationary
and SUSY scales below.  For the moment we note only that the simplest
resolution of the cosmological moduli problem\bkn\casas is to give the moduli
mass at a scale higher than the SUSY breaking scale.  If moduli masses
arise from nonperturbative SUSY breaking then they typically dominate
the energy density of the universe until it is too low for
nucleosynthesis to occur.  This can be avoided if moduli masses are
generated at a higher scale. If the moduli are the
inflatons, the high scale would then be the natural vacuum energy scale
during inflation.  Finally we note, for what it is worth, the coincidence
between the vacuum energy scale ``determined by COBE'' and the putative scale
of coupling unification in SUSY GUTS
\ref\susyunif{U.~Amaldi et al.,  {\it Phys. Rev.} {\bf D36},(1987),1385;
  G.~Costa et al., {\it Nucl. Phys.}{\bf B297},\break (1988),244;
  P.~Langacker, M.-X.~Luo, {\it Phys. Rev.}{\bf D44},(1991),817;
  U.~Amaldi, W.~de Boer, H.~F\"urstenau, {\it Phys. Lett.}{\bf
B260},(1991),447;
  J.~Ellis, S.~Kelley, D.V.~Nanopoulos, {\it Phys. Lett.}{\bf
B260},(1991),131;
  F.~Anselmo, L.~Cifarelli, A.~Peterman,A.~Zichichi, {\it Nuovo Cimento}{\bf
104A},(1991),1817;
  W.J.~Marciano, {\it Ann. Rev. Nucl. Part. Phys.}{\bf 41},(1992),469.}.

In order to begin our study of modular cosmology, we must deal with the problem
of initial conditions.
 In principle, one would want to give a completely quantum mechanical
and string theoretic description of initial conditions.  Unfortunately, neither
classical nor quantum string theory is sufficiently well developed to make
a really fundamental attack on this problem\foot{See however, the very
interesting
contributions of \ref\tseytvafa{A.Tseytlin, C.Vafa, {\it Nucl.
Phys.}{\bf B372},(1992),443.}.}.  We will therefore follow tradition and
assume that at an energy scale just below the Planck scale, the conditions
in at least some small patch of the universe, can be described by the
semiclassical dynamics of moduli fields coupled to gravity.
We argue that the lagrangian describing the moduli is a nonlinear model on
a noncompact target space of finite volume.
 The potential term in the lagrangian
is of order $(10^{16} - 10^{17}\quad GeV)^4$ or smaller.
We will show that this
means that the horizon volume becomes highly inhomogeneous by
the time the energy density
falls to the scale of the potential.  Heuristically this occurs because,
at energy densities much higher than the potential, the
kinetic energy of the homogeneous modes of the fields redshifts like $R^{-6}$,
while the energy in inhomogeneous fluctuations red shifts only like $R^{-4}$.
Consequently, when the energy
 density falls to the scale of the potential, different
domains of space will fall into different local minima of the potential.

We next observe that string theory moduli provide two generic kinds of
domain wall excitations \foot{Domain walls in the low energy theory of
moduli have previously
been considered by M.Cvetic and collaborators \ref\cvetic{
 M.Cvetic, F. Quevedo, S.J. Rey
{\it  Phys.Rev.Lett.}{\bf 67},(1991),1836;
{\it TOPOLOGICAL DEFECTS IN THE MODULI SECTOR OF STRING
THEORY.}
M. Cvetic, Presented at Strings
'91 Workshop, Stony Brook, N.Y., May 20-25, 1991.
Published in Strings: Stony Brook 1991:29-35,
hep-th@xxx.lanl.gov - 9109044 ;
M.Cvetic, S. Griffies, S.J. Rey,
{\it Nucl.Phys.}{\bf B381},(1992),301,
hep-th@xxx.lanl.gov - 9201007 ;
{\it STRINGY DOMAIN WALLS AND OTHER STRINGY TOPOLOGICAL
DEFECTS.}
M.Cvetic, Presented at
Summer School on Particle Physics, Trieste, Italy, Jun 15 - Jul 30, 1991.
Published in Trieste HEP Cosmol.1991:1027
hep-th@xxx.lanl.gov - 9202067 ;
M. Cvetic, R.L. Davis, {\it Phys.Lett.}{\bf B296},(1992),
hep-th@xxx.lanl.gov - 9205060
M.Cvetic,{\it  Phys.Rev.Lett.}{\bf 71},(1993),815,
 hep-th@xxx.lanl.gov - 9304062.}}.
First note that
the resolution of the Dine-Seiberg problem of string theory
requires the existence of a ridge in the moduli space potential,
which separates
the true finite coupling vacuum from the weak coupling region.  This
implies that there are
 quasistable domain wall configurations, in which the fields
traverse the ridge as a function of one spatial coordinate.  Secondly,
moduli space appears to contain noncontractible loops.  A configuration
in which the moduli traverse one of these loops, which goes through
the
minimum of the effective potential, as a function of
one spatial coordinate,
would be a topologically stable domain wall in Minkowski space.
There is another kind of generic domain wall associated with the
discrete modular symmetries of string theory.
Our observations
about the nature of the initial state suggest that many domain walls
of all types will
be produced in the course of the expansion of the universe.

When gravity is
taken into account, we find that, {\it if the potential at the top of the
domain wall is flat enough}, the center of each domain wall inflates
forever\ref\vilenkin{R.Basu, A.Vilenkin, {\it Phys. Rev.}{\bf
D46},(1992),2345; {\it Phys. Rev.}{\bf D50},
(1994),7150; \break A.Vilenkin, {\it Phys. Rev. Lett.}{\bf
72},(1994),3137.
}\ref\linde{A.Linde, {\it Phys. Lett.}{\bf 327B},(1994),208; A.Linde,
D.Linde, {\it Phys. Rev.}{\bf D50},(1994),2456.}.
Thus, by the time the average energy density
of the initial patch has fallen significantly below the height of the ridge,
most of the volume of the patch will be covered by inflationary domains
which originated as domain walls.  Quantum fluctuations will drive
subvolumes of these inflationary domains to \lq\lq
roll off the potential" towards
a nearby minimum of the potential.  In the case of walls draped over the
ridge between weak and strong coupling the fluctuation will move in
the direction of
either the weak or finite coupling regions with about equal probability.
Thus the problem exposed in \problems{} is substantially mitigated.  In this
scenario, fifty percent of the region that undergoes inflation, eventually
settles in to the correct vacuum state at finite coupling.  This is at least
a partial solution of the problem raised in \problems .

In section 4 we explore the discrepancy between the vacuum energy scales
required by COBE and by the SUSY solution to the hierarchy problem.
One possible explanation of this discrepancy
 exploits the fact that the potential depends
exponentially on the inverse of the unified fine structure constant.
Thus, if the point on the ridge
separating weak and strong coupling is at a larger value of the coupling
than the true minimum, the potential on the ridge could be many orders of
magnitude larger than the square of the F term at the minimum.  This by
itself could be the explanation of
 the discrepancy between the COBE and SUSY scales. We will refer to this
 as the \lq\lq one component'' scenario for explaining the discrepancy
 in scales.

In order to solve the problem of moduli, a more complicated mechanism
might be necessary.  For example, both \bkn and \casas point out
that one way to solve the problem of modular domination of the energy density
of the universe, is to assume that moduli get their masses from nonperturbative
SUSY preserving dynamics at a very high scale.  This scale might be identified
with the high vacuum energy scale required by the COBE observations.
The superpotential would then be the sum of two pieces, the first of which
gives rise to inflation and the second of which breaks SUSY.  The discrepancy
in scales is explained by assuming that these two pieces are proportional
to different exponentials of the inverse string coupling.
We call this the {\it two component scenario} for modular cosmology.

In a previous paper\ref\papertwo{T.Banks, M.Berkooz, P.J.Steinhardt,
{\it The Cosmological Moduli Problem, Supersymmetry Breaking, and
Stability in Postinflationary Cosmology}, Rutgers preprint RU - 94 - 92,
hep-th@xxx.lanl.gov 9501053.}
 some of the authors have investigated this
scenario in some detail and found that it is very difficult to realize.
The required SUSY preserving dynamics seems to realized, if at all, at
special points in moduli space where the number of light chiral multiplets
charged under the hidden sector gauge group, undergoes an increase.  There are
no known points in moduli space with this property.  Even if one could be
found, this mechanism cannot given rise to a dilaton mass
unless we allow cancellations of two field theoretic effects of different
nominal orders in the weak coupling expansion (as in the racetrack models).
Despite these difficulties, one might still wish to explore the idea
that the COBE data are really telling us about a new scale in physics,
associated with the vacuum of some strongly coupled SUSY gauge theory.
The rough coincidence of the required scale with the ``observed'' unification
scale lends impetus to speculations in this direction.

In the modular scenario, inflation does not occur on the ridge between
strong and weak coupling, which is of negligibly small height during the
inflationary era.  However we show that there are domain wall solutions
associated with noncontractible loops in moduli space which can drive
defect inflation.

The problem of \problems takes on very different aspects in
the one and two component proposals for explaining the difference between
inflationary
and SUSY scales.  Indeed, in the first proposal, inflation takes place
at a much higher energy scale than the smallest barrier between the true
vacuum and the weak coupling region.  The arguments of \problems
would lead us to expect that the system will run over the low barrier
into the weak coupling region.  Thus our proposal for solving the
dilaton runaway problem may not work if we also attempt to explain
the difference between inflation and SUSY scales in terms of a
superpotential with a single exponential\foot{
However, we have also made
some observations which suggest that the dilaton
runaway problem may not be as
severe as was envisaged in \problems .
We present this analysis in an appendix. }.

If the discrepancy in scales is explained by a superpotential which is the
sum of two exponentials then the dilaton does not run to infinity.
During the inflationary era, and the period before the moduli fields
settle in to their vacuum values, the dilaton feels a large potential
which may be assumed to confine it to the vicinity of a point$S_0 (M)$
($M$ are the moduli fields) which traces out a trajectory on the space
of nondilatonic moduli (Fig. 1).
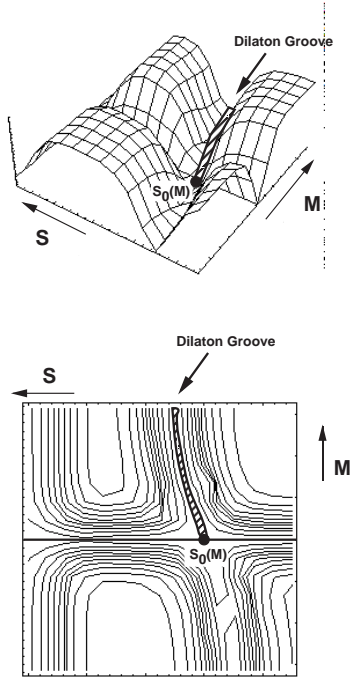
\ifig\fone{In the two component scenario for Modular Cosmology, the dilaton
is trapped in a \lq\lq groove'' $S_0 (M)$ until the moduli/inflaton fields
return to their minimum.  If the end of this groove is on the strong coupling
side of the true dilaton minimum, the dilaton does not run into the weak
coupling region.}
{\epsfysize=8cm \epsfbox{fig3.draw.epsf}}
This potential vanishes as $M\rightarrow 0$.
When the moduli fields are small enough, the energy density in the
large part of the potential, falls to the level of the smaller part of the
potential.  If, as $M\rightarrow 0$,
 $S_0 (M) $ is smaller than (i.e. lies to the strong coupling
side of) the true minimum of the small potential,
then the dilaton will always end up at the true minimum.

Thus, the combined requirements of stopping the runaway dilaton, and
of explaining the discrepancy between the inflationary and SUSY scales
for the vacuum energy,
strongly suggest that the two component approach to modular cosmology is
the correct one.

The modular cosmology outlined in section 5 is of course highly speculative.
Moreover, it suffers from the Postmodern Polonyi Problem (PPP)outlined in
\bkn\casas.  Even in the scenario in which geometric moduli obtain a
mass much larger than the weak scale, the dilaton dominates the energy
density of the universe until it is too low for nucleosynthesis
to occur.
In a previous
paper\papertwo, some of us have explored possible ways of solving the PPP.
At present, there are no completely satisfactory resolutions
 of this problem.

We summarize what we have learned in the concluding section, 6.
We point out that many of the observations that we have made are valid
in a more general context than superstring inspired inflationary models.
In particular, the need for fields whose natural range of variation
is the Planck scale, appears quite general.  This requirement more or less
follows from field theoretic naturalness.  The necessity of introducing
topological defects as the seeds of inflation comes from
the fact that observational constraints on relic gravitons and on the
microwave background fluctuations tell us that the energy density during
inflation was ten orders of magnitude (or more) below the Planck
density.  In many models, generic initial conditions starting at the Planck era
do not
lead to the degree of homogeneity required in conventional inflation
models, at the inflationary energy scale.  Defect inflation is a very general
resolution of this problem.  However, in models where inflation takes
place
in unbounded regions of field space, or from a metastable minimum, we
can have inflation without introducing defects.  Thus defect inflation
is more general than superstring inspired models but not a universal
requirement.

Note Added: After this paper was written, we were informed by S. Thomas
that he had explored ideas similar to those discussed here.

\newsec{Before Inflation}

To be frank, we are as ignorant as everyone else of the proper initial
conditions for cosmology.  Undoubtedly the full story
involves quantum mechanics, and the correct short distance theory of geometry,
which might be string theory.
Conventional accounts assume that the universe we observe arose from a much
smaller universe, or a small patch, henceforth called
the inflationary patch, in a meta-universe we may never
see.  The size of this patch is taken to be sufficiently large that quantum
mechanics of the geometry of spacetime, and the details of the correct
short distance theory, may be ignored.  Clearly such an assumption may
be wrong, but as always, one does what one can or does nothing at all.
A more fundamental approach to stringy cosmology has been studied in
\tseytvafa .

There is one fairly generic consequence of the assumption that the
inflationary patch is much larger than the string scale, namely that the
homogeneous modes of scalar fields in
the inflationary patch may be described by classical dynamics.  Indeed, their
lagrangian has the form
\eqn\modlag{L =R^3 (t)(K_{ij}[\phi ]{\dot\phi^i}{\dot\phi^j} - V(\phi ))}
The volume $R^3$ of the patch, is by assumption large compared to the Planck
scale.  As a consequence, the stationary phase
approximations for the dynamics of the zero modes is always valid.  The initial
conditions for these classical degrees of
freedom are presumably determined only probabilistically, by some quantum
process at shorter distances.  For our purposes,
we will need to make only a few general assumptions about this probability
distribution.

The initial quantum state of the rest of the degrees of freedom is something of
a mystery.  Conventionally it is assumed to
be such that it does not make significant contributions to the equation for the
coherent classical gravitational field.  We will make this assumption
initially, but we will see that the dynamics of moduli is such that it
rapidly becomes untenable.

Now consider the classical lagrangian of the moduli.  The space of fields is a
complex manifold, equipped with a Kahler metric.
To all orders in perturbation theory the low energy effective lagrangian has
the form
\eqn\modlagtwo{\sqrt{-g}[\partial_{\mu} m^i g^{\mu\nu}
\partial_{\nu} {\bar{m}^{\bar{j}}} K_{i\bar{j}}] }
where $K$ is the Kahler potential of the moduli.

We will include the dilaton axion supermultiplet
as the first of the moduli fields, $m^1$.  However, we will often
single out this field and call it $S$, following the usual convention in
the literature.   To all orders is perturbation theory
$K$ depends only on $m^1 + {\bar{m}^{\bar{1}}}$ .
We should also note that the mass scale in $K$ is the string scale
$M_S = g_S M_P = g_S{} 10^{19}$ $GeV$.
where $g_S$ is the unified string coupling, possibly of order
 $1\over \sqrt{2}$.  However, we will not have occasion to do
 calculations accurate enough
to distinguish $M_S$ from $M_P$, so we will usually refer to them both
as $M_P$.

Usually, in writing down an effective field theory
at a scale well below the Planck mass, we keep only the first few terms in an
expansion of the lagrangian in inverse powers of the Planck mass
, and forbid the discussion of field values of order the Planck mass.  This
is not the correct procedure for moduli in string theory.
We know many things about how the lagrangian changes when the moduli change
by amounts of order the Planck mass.  Indeed, this is the
natural scale of variation for these fields.  We can compute from the
underlying string theory for example, just what happens to
the Yukawa couplings of quarks and leptons when we change the size of the
internal manifold by an amount of order the Planck scale.
Thus the power counting in the effective field theory of moduli is that we
should keep only terms with small numbers of derivatives
but arbitrary dependence on the moduli fields.  The situation is somewhat
similar to that of pions in the effective chiral lagrangian of
QCD.  We do not expand this out in powers of the canonically normalized pion
field, because symmetry considerations tell us how the
lagrangian depends on low momentum pion fields, even when the fields are
of order $4\pi f_{\pi}$.  In the string case, the field space
is noncompact, and it is explicit computations rather than symmetry that fix
the field dependence.

An important property of the lagrangian for moduli is that the volume of
moduli space determined by the metric in the lagrangian, is finite
\moorehorne {}, despite
the existence of noncompact regions.  This appears to be true classically
for all
noncompact regions of moduli space which are currently well understood.
For some of these regions, we cannot be sure that quantum corrections
do not change the metric.  However, the noncompact region which will
be of primary interest to us in this paper is the extreme weak coupling
region.  Here we can rely on classical considerations.

The finite volume of moduli space is an important constraint on the
choice of \lq\lq generic'' initial conditions.  If moduli space really
had infinite volume, then one would have expected a generic initial
condition to be somewhere deep in the noncompact region of field space.
(Note that general arguments suggest that the potential will vanish in
many of these noncompact directions).  In particular, we would expect to
find the initial value of the dilaton deep in the weak coupling region.
It would then be highly unlikely for the universe to evolve to a finite
value of the string coupling.

The authors of \moorehorne argued that
the
dynamics of the zero modes on moduli space was chaotic, implying that
after a few Lyapunoff times any reasonable distribution of initial
conditions gets spread uniformly (in the finite volume measure) over
moduli space.  Unfortunately, as we will show below, it is not a good
approximation to restrict attention to the zero modes.  Inhomogeneous
modes grow and the coupling of the zero mode to the inhomogeneous modes
damps the chaotic motion on times scales shorter than the Lyapunoff
time. Thus, the dynamical mechanism proposed in [12] cannot effectively
distribute the initial conditions uniformly in the finite
volume measure of moduli space.
Nevertheless, we will make the assumption that the distribution
of initial conditions suggested by that argument --- that
noncompact regions still have finite volume --- is valid for
modular cosmology.

The second fact that will be important for our considerations is that
the scale of the potential in regions of interest is far below the
Planck scale.  This is a phenomenological input to our calculations.
We are trying to construct an inflationary model, and in particular
to preserve the standard inflationary prediction of fluctuations
in the microwave background.  This requires us to use a potential
which is ten to twelve orders of magnitude smaller than the Planck
scale in the region where inflation takes place.  Near its minimum,
the potential will drop nearly to zero, corresponding to a state with
very small cosmological constant.

Consequently, if we begin our considerations of modular dynamics
at a time when the energy density is just below the Planck scale,
then we can neglect the potential initially.
It is possible to solve for the reaction of a Robertson-Walker
metric to the energy density of the homogeneous modes of a general
nonlinear model with no potential.  Indeed, because of the
simplicity of the nonlinear lagrangian, the pressure and energy
density terms in the stress tensor, satisfy the exact equation of
state $p = \rho$.  It follows that $\rho \propto R^{-6}$ ,
and that $R \sim t^{1\over 3}$.  The expansion is subluminal
and the energy density falls off extremely rapidly.

Now consider corrections to this behavior coming from inhomogeneous
modes of the field.  The equation for the zero mode is
\eqn\zeromode{{\ddot{\phi}}^l + \Gamma^l_{is} {\dot{\phi}}^i
{\dot{\phi}}^s + 3 H{\dot{\phi}}^l = 0,}
where $\Gamma^i_{jk}$ is the Christoffel connection on moduli space.
This is a geodesic equation with a friction term.   The trajectory
is a geodesic, but it is followed at a different rate than it would be
in a flat spacetime.  If we make a Fourier decomposition of the
deviations of the field from this trajectory,
$\delta\phi^l (x,t) = \delta\phi^l (t,{\bf k}) e^{i{\bf k x}}$,
then the Fourier components satisfy the linearized equation
\eqn\lin{{\ddot{\delta}}\phi^l + 2\Gamma^l_{is}{\dot{\phi}}^i
\dot{\delta}{\phi}^s + \Gamma^l_{is,q}{\dot{\phi}}^i{\dot{\phi}}^s
\dot{\delta}{\phi}^q + 3H \dot{\delta}{\phi}^q + {k^2\over
R^2}\delta\phi^l = 0.}

Now choose coordinates on moduli space for which the initial geodesic is
one of the coordinate lines , $\phi^0 (t)$, and such that $\Gamma^l_{00}$
vanishes along the geodesic.  As we explained above,
$H = {1\over {3(t + t_0)}}$, and $R(t) \sim t^{1\over 3}$.  This implies
that $\phi^0 (t)\propto (t + t_0)^{-3}$.  The equation for
$\delta\phi^0$ is particularly simple
in these coordinates, and does not involve the other components of
the deviation from homogeneity.  It is
\eqn\linpar{ {\ddot{\delta}}\phi^0 +{1\over (t + t_0)}\dot{\delta}\phi^0
+ {k^2 \over R^2}\delta\phi^0 = 0}
For the given form of $R(t)$, this is a variant of Bessel's equation.
The large time asymptotics of the solution are
\eqn\asymptot{\delta\phi^0 \sim t^{-{1\over 3}} F(t^{2/3} )}
where $F$ is a trigonometric function.  This leads to an energy density
that scales like $t^{-{4\over 3}}$ or $R^{-4}$.  Thus the homogeneous
energy density falls relative to the inhomogeneous component by two powers
of $R$.

The equations of modes orthogonal to $\phi^0$ behave in a similar
manner or worse.  These equations contain the sectional curvatures
of moduli space, which are frequently negative.  This leads to
a further growth in the amplitude of the inhomogeneous modes relative
to the homogeneous one.

We see that by the time the energy density has fallen to the scale
of the potential, our initial assumption of homogeneity has lost
all plausibility.  We have just shown that within each original
horizon volume the inhomogeneous energy density has grown relative
to the homogeneous component by a factor of $10^{10/3}$.
Furthermore, the expansion is subluminal, so the current horizon
volume contains many initial horizon volumes.  The current size
of an initial, Planck, horizon volume is about $10^5$ Planck
volumes, while the current horizon volume is $10^{15}$ Planck
volumes.
 Conventional models
of inflation assume the existence of a homogeneous region at least
as large as the current horizon volume at a time when the energy
density is dominated by the potential.  In the present context
the existence of such a region seems dubious.

\newsec{\bf Defect Inflation}

All is not lost.  Once the potential has come to dominate the
energy density, fields tend to fall into its local minima.
Generic features of string theory lagrangians, and of moduli space,
lead to the existence of a complicated potential surface which supports
topological defects.  Thus string theory
provides a natural setting for {\it defect inflation},
an idea first proposed by Linde \linde{} and Vilenkin \vilenkin{}.

As an example of the kind of defect that must be present in string
theory, let us assume the existence of a solution
of the Dine-Seiberg problem. Then we know that the potential has
at least two local minima, the \lq\lq physical vacuum'' and the
\lq\lq weak coupling region'' , separated by a ridge in moduli space.

Consider now two adjacent regions of order the horizon size
at the time the potential comes to dominate the energy density.
Since {\it the total volume of moduli space is finite} \moorehorne {}, there is
a reasonable probability that one of the regions will be evolving towards
 weak coupling, while the other evolves towards the physical
vacuum.  This will lead to the formation of a domain wall.
Thus, we claim that generic initial conditions for moduli near the
Planck scale, lead, as a consequence of our mild assumptions
about the modular potential, to the formation of many domain walls
straddling the ridge between the weak coupling region and the physical
vacuum.

In fact, it is likely that string theory contains other generic kinds
of domain wall solution.  Indeed it appears that the moduli space of
heterotic string vacua has noncontractible loops.  There is a subspace
of moduli space called the conifold subvariety\ref\conifold{P.Candelas,
P.Green, T. Hubsch, {\it Nucl. Phys.}{\bf B330},(1990),49; P.Candelas,
X.C. de la Ossa, {\it Nucl. Phys.}{\bf B342},(1990),246.} which has
complex codimension one.  At these points in moduli space, physical
couplings blow up.  It is not clear how one should interpret these
singularities, but certainly one possibility is that one must excise
these singular points from the space of classical solutions.  In that
case the space would have noncontractible loops.
 Assume that the potential for moduli
has an isolated minimum on moduli space and consider a noncontractible
loop
going through the minimum.  A field configuration in which the moduli
traverse the noncontractible loop as a function of one spatial
coordinate, asymptoting to the minimum of the potential as the spatial
coordinate goes to infinity,
will be topologically stable.  The minimum energy configuration with
this topology will be a domain wall\foot{If there was no potential on
moduli space, the defect would be a cosmic string, but in the
presence of a potential, the string has a domain wall attached to it.}

In addition to these potential noncontractible loops, the moduli space
of string solutions certainly has orbifold points.
For example, at tree level the moduli space of a toroidal compactification
is two copies of the the fundamental domain of $SL(2,Z)$ in the upper
half plane with Poincare metric, which has the usual orbifold points.
An orbifold space (i.e. a generalized cone) can be viewed as a covering
space, with points which are related by some discrete group of symmetries,
identified.  Now consider a potential on the orbifold which has a
minimum at a point not invariant under the group of identifications.
On the covering space the potential will have two degenerate minima, and
there will be a topologically
stable domain wall configuration of the field theory on
the covering space.  Considered as a field configuration on the
orbifold, this domain wall defines a state in the {\it twisted sector}
of the quantum field theory whose target space is the orbifold.
It will be a locally stable domain wall, since the local stability
analysis will be the same in the covering and the orbifold theories.
In the orbifold theory
there is only one minimum of the potential, and the domain wall
configuration describes a closed loop in the moduli space encircling the
orbifold singularity.  The existence and local stability of the
domain wall are guaranteed by the analysis of the symmetries of the
potential on the covering space.

In the orbifold theory, the domain wall is no longer topologically
stable. Its nonperturbative instability can be described in two ways.
In orbifold language the domain wall corresponds to a closed loop in
target space which surrounds the orbifold point.  We can ``slip it over
the tip of the cone'' in a finite region of space, and contract the
loop.  Alternatively we can
think of the orbifold theory as a discrete gauge theory on the covering
space. In the absence of a potential it will have cosmic string
configurations with string tension determined by the Planck scale.  Once
the potential is introduced, these strings are the boundaries of domain
walls, precisely the walls we have been talking about.  The gauge
invariant description of ``slipping the loop over
the tip of the cone'' is equivalent in gauge variant variables to the
nucleation of a loop of cosmic string in the domain wall, which will
expand and destroy the wall.

The cosmic string solution exists even without the generation of a
potential by nonperturbative effects.  Thus we would imagine that
its core energy is of order $M_P$. Since the string core energy is much
larger than the surface tension of the wall, the wall will be highly
metastable\foot{Formally, the classical gauge
field kinetic term of an orbifold gauge theory vanishes (the orbifold is
the extreme strong coupling limit).  In a supersymmetric theory we might
worry that the cosmic string had zero energy to all orders in
perturbation theory. However, the local stability of the domain wall,
certainly means that the cosmic string instability is a nonperturbative
effect. The potential gives the string a core energy even if it did not
have one in perturbation theory.
The estimates of instanton actions in the
field theory of moduli made in \papertwo suggest that the walls would
have enormously long lifetimes even if the string core energy were of
the same order as the surface tension. Furthermore, as we will see
below, the domain wall cores inflate (if the wall is sufficiently
smooth) when coupled to gravity, and all of these flat space
instabilities become irrelevant, as long as the flat space lifetime is
much longer than the e-folding time.}.  Note that global
cosmic strings associated
with truly noncontractible loops in moduli space will provide a similar
mechanism for the decay of the domain walls which traverse these loops.
The similarity between the dynamics of these two types of domain walls
leads us to abuse language and refer to both types of walls as
being related to noncontractible loops, even though
the loop on the orbifold is topologically contractible.

  Kibble's argument\ref\kibble{T.W.B.Kibble, {\it J. Phys.}{\bf
  A9},(1976),1387.}
and our observations about the number of initially
causally disconnected regions in a horizon volume at the time the potential
becomes important, suggest that many domain walls associated with both
types of \lq\lq
noncontractible loop'' in moduli space will be
formed as the energy density falls below the scale of the potential.

In order to proceed, we will have to make two more assumptions
about the potential on moduli space.  The first is, we believe,
rather generic.  We assume that the ridge between weak and strong
coupling has a saddle, with exactly
one unstable direction (namely the one leading down to the two minima).
We will further assume a certain degree of flatness
of the maximum at the top of the ridge (the precise conditions
will be stated below).  Alternatively, we will assume
that the potential near the top (the point of highest potential energy)
of the noncontractible domain wall is sufficiently flat.
We will show below, that the
assumption of flatness leads to {\it defect inflation}.

 The first requirement for inflation is
an effective potential with a
``sufficiently  flat" segment.
In the flat segment, the equation of motion for
 the inflaton field, $\phi$, must be well-approximated
by the slow-roll equations:\ref\slowroll{A.Linde, {\it Phys. Lett.}{\bf
108B},(1982),389; A.Albrecht, P.J.Steinhardt, {\it Phys. Rev. Lett.}
{\bf 48},(1982),1220.}
\eqn\slowrollone{
3 H \dot{\phi} = - V'(\phi),}
and
\eqn\slowrolltwo{
H^2 = ({\dot{a}\over a})^2 = {8 \pi G V(\phi)\over 3},
}
where $a$ is the Robertson-Walker scale factor, $G=M_p^{-2}$ is
Newton's constant, and $V(\phi)$
is the effective potential.  The slow-roll approximation is
 valid for a span of
the potential over which $\ddot{\phi}/3 H \dot{\phi}$ and
${1\over2} \dot{\phi}^2/V(\phi)$
are negligible.  This requires that the flat segment of the effective
potential satisfy:\ref\potconds{P.J.Steinhardt, M.S.Turner,
{\it  Phys. Rev.}{\bf D29},(1984),2162.}
 \eqn\conda{
({V'\over  V})^2 \ll 48 \pi G = 48 \pi M_p^{-2}
}
and
\eqn\condb{
{1\over 4} ({V'\over  V})^2 -
                {1 \over 2} ({V''\over  V})
\ll 12 \pi G = 12 \pi  M_p^{-2}.}

A second condition is that
 the slow-roll condition must be satisfied for a sufficiently
long stretch that the universe undergoes 60 or more e-folds of inflation
as $\phi$ rolls down the potential.  The reasoning for defect inflation
is slightly different than for other forms of inflation, although
the quantitative
constraint is the same in the end: We envisage a situation in which
the effective potential has two degenerate minima (possibly identified
by a discrete gauge symmetry) separated by a ridge.
This allows the possibility for
a locally stable domain wall in which $\phi$ ranges from
one minimum in one region of space to the other minimum in another region.
Somewhere between, $\phi$ must traverse the top of the ridge of the
effective potential, $\phi  \approx \bar{\phi}$.
  If the ridge near $\phi  \approx \bar{\phi}$ is sufficiently
flat (satisfies the conditions above), the defect core itself begins
to inflate.  The core is stretched but, at the same time,
quantum (de Sitter) fluctuations will drive $\phi$ away
from the top of the ridge in some regions.  The random fluctuations will
drive some regions towards $\phi>\bar{\phi}$ and others towards
$\phi<\bar{\phi}$. The radius of any such region initially will be
of order a Hubble horizon radius.  After one e-folding, the radius
is stretched but, at the same time, subsequent quantum fluctuations
will have split the region into subregions (each with Hubble horizon
radius) with different values of $\phi$.  In particular, a region
with $\phi>\bar{\phi}$ at first will be subdivided into subregions
which will include ones with $\phi<\bar{\phi}$.  This subdivision
process will end in a given region when $\phi$ grows large enough
that the classical evolution of $\phi$ dominates the quantum fluctuation
effect; let us call this crossover point $\phi= \phi^{*}$.
At this point, a typical homogeneous region still has radius of
order $H^{-1}$.  For inflation to explain why the present observable
universe is so homogeneous, the final volume of the homogeneous
regions needs to grow by 60 e-folds  or more.
Consequently, we require at least 60 e-folds of inflation during
the classical slow-roll process where $\phi>\phi^{*}$.
This constraint is, roughly: \potconds

\eqn\condd{
{4 \over 9} ({-V'' \over   H^{2}}) \ll {1 \over N_{e}}
}
where $N_e \approx 60$ is the minimum number of e-folds to solve
the homogeneity problem; or, equivalently,

\eqn\condc{
 ({-V'' \over   V}) \ll {6 \pi \over N_{e}}
}
These conditions, Eqs. (\conda) through (\condc),
 are necessary and sufficient to achieve the minimal
number of e-foldings of inflation to resolve the flatness, horizon and
monopole problems.

Let us consider what happens if the ridge between strong and weak
coupling, or the potential at the top of a noncontractible domain wall,
is sufficiently flat
to satisfy
the slow-roll conditions.  Without loss of generality, we can expand
the
effective potential about a point near the top of the ridge, $\phi=
\bar{\phi}$
as:
\eqn\taylor{
 V \approx V_0 (1 - \alpha^2 {(\phi-\bar{\phi})^2\over  M_p^2})
}
In flat space, the wall thickness would be equal to the curvature of
the effective potential,
$ \delta \sim \alpha (V_0/M_p^2)^{{\ha}}$.
The Hubble parameter in the interior of the wall is $H \approx
(8 \pi G V_0 /3)^{{\ha}}$.  If $\delta \ll H^{-1}$, gravitational
effects are negligible.  However, if $\delta > H^{-1}$, the region of false
vacuum with  $\phi$ near the top of the ridge and $V\approx V_0$ extends
over a
region greater than a Hubble volume.  Since the top of the  ridge satisfies
the conditions for inflation, the interior of the wall inflates.

Once started, defect inflation never ends.
  Although $\phi$ in regions of the
defect core may roll down the potential and those regions may reheat into
a Friedmann-Robertson-Walker-like patch similar to our own Hubble volume,
 topological
constraints require that there always remain some core region with $\phi
\sim
\bar{\phi}$.  Due to inflation, the core thickness grows exponentially
 with time.
In fact,
the inflation is so rapid that the defect never
settles into a stable, minimum energy configuration.

The condition for defect interiors to inflate, $\delta > H^{-1}$,
corresponds to
\eqn\inflacond{
  \alpha^2 <  8 \pi/3 .
}
In this regime,
\eqn\ques{
  ({V'\over V})^2 \sim {\alpha^4 (\phi-\bar{\phi}^2) \over  M_p^4},}
which satisfies the first inflationary condition, Eq. (\conda), for
\eqn\dques{
 \Delta \phi \equiv (\phi-\bar{\phi}) \ll {\sqrt{48 \pi} \over
\alpha^2} M_p.}
The second inflationary condition, Eq. (\condb), reduces to:
\eqn\secd{
 \Delta \phi \ll  \left[ {12 \pi \over \alpha^4} -
{2 \over \alpha^4}\right]^{1/2} \, M_p}
The expression in  square brackets is positive for all $\alpha$
satisfying the third inflationary criterion (using $N_e =60$),
Eq. (\condc):
\eqn\thrd{\alpha^2 < \pi/20}
(This last conditions is the most stringent condition on $\alpha$.
In deriving these relations, we have assumed that
 $\alpha^2 (\Delta \phi/M_p)^2 \ll 1$, which is consistent with
the previous constraints on $\alpha$ and $\Delta \phi$.

It is interesting to note that the conditions for inflating domain wall
cores do not require fine tuning of the dimensionless parameter $\alpha$
in the potential.  This is a consequence of our choice of a \lq\lq
natural'' potential with Planck scale variation of the fields.  Models
of defects arising at the Grand Unification scale, would require fine
tuning of dimensionless parameters.  We may however have been a bit too
optimistic in the above estimates.  We have used the conventional
${8\pi\over 3}$ normalization of Einstein's equations, appropriate for
a scalar field with canonical kinetic term.
In string theory, the
moduli fields have kinetic terms whose normalization is related to
Einstein's action.  At tree level, the conventional factor of
${8\pi\over 3}$ is simply
not there.  In the philosophy of \banksdine , the actual kinetic terms
of the scalar fields are at present uncomputable, so it is not clear
what the proper normalization is.
Perhaps it would be more conservative to assume the tree level
normalization (though not the tree level form of the Kahler potential).
In that case we would replace every $\pi$ in the formulae above by
${3\over 8}$.  Note that the inflationary conditions are still satisfied
for $\alpha^2$ of order $.02$.  We take this as evidence that in modular
cosmology defect inflation requires at most only a rather mild
fine tuning.

The motivation for invoking defect-cores as the seeds of inflation is to
exponentially enhance the {\it  a posteriori} probability that a region equal
to our present Hubble horizon emerged from a patch of space-time that once
underwent inflation.  There are three enhancement factors.  First,
beginning from
random initial conditions,
the probability of forming defects with $\phi \sim \bar{\phi}$ at the
core
is exponentially higher than the  probability of forming some
bounded region (with no topological stability) with $\phi \sim \bar{\phi}$.
  The
probability of forming the bounded region scales according to the ratio of
 the flat
portion of the potential to the entire potential surface, which is small.
The probability of forming a defect scales according to the areas of the
basins on the potential surface which draw $\phi$ towards the degenerate
minima, which can be  exponentially greater.  The second enhancement factor
 is that
the inflation is eternal.  Almost all models of inflation have an eternal
character:  although some regions slow-roll down the effective potential,
 random
quantum fluctuations kick the value of $\phi$ in some regions back towards
the top of the potential where they continue to inflate.   The  defect
cores exhibit similar behavior. However, on top of that, there is a classical
contribution to  eternal inflation due to the classical, topological
constraint
that guarantees some region with $\phi \sim \bar{\phi}$ even if ${h}
\rightarrow
0$.
Finally, we have noted that the natural initial conditions for modular
cosmology do not lead to homogeneity over a horizon size when the energy
density is near a maximum of the potential consistent with the
proper amplitude for primordial fluctuations.  By contrast, if the
potential is flat enough, the defect is homogeneous over a horizon
volume.

A technical issue in applying these concepts in a superstring model is
that the
height and  curvature
 ($\alpha (V_0/M_p^2)^{\ha}$)
of the effective potential at the top of the ridge is  not
uniform along the ridge.  (It is difficult, if not impossible,
 to construct a model in
which the ridge satisfies the conditions for inflation everywhere along
the ridge.)
Consequently, a field configuration  which traverses the
ridge may contain a
core which, initially, does not satisfy the conditions for inflation.
Note, though,
that configurations connecting the vacua and crossing at different passes
along the ridge have different energies.
It should suffice, though, if  the minimal energy field
configuration  connecting the two vacua traverses the ridge at that
segment where inflationary conditions are satisfied.  Then, any initial
field configuration connecting the vacua will settle into one in which
the core inflates.  Or, there may be ripples  (with saddle points) along
the ridge.  It suffices if one or more of the saddle points satisfy the
inflationary conditions.  Depending on the depths of the
saddle points,
all or  most of topological defect cores will relax into an inflationary
saddle point.

\newsec{\bf Solutions and Problems}

\subsec{The Fifty Percent Solution}

In the case of defects straddling the ridge between strong and weak
coupling, the scenario of defect inflation that we have just described, leads
to what one may call the {\it fifty percent solution}
of the cosmological dilaton problem discovered in \problems .
In such a situation, postinflationary history always starts with
the initial values of fields perched on the ridge between the
weak coupling and physical vacua.  The direction in which the fields
fall off the saddle is determined by quantum fluctuations, {\it i.e.}
it is essentially random.  Thus, with fifty percent probability
the universe rolls toward the correct vacuum state. If the inflationary
ridge is not much higher than the lowest ridge separating
the vacuum from the weak coupling region, the universe
will surely end up in the correct vacuum state.
 We have emphasized
that our scenario leads to the formation of many domain walls,
so the quantum ensemble is realized in terms of an ensemble of different
inflationary regions spread through the universe.

It seems perfectly acceptable to us to have a theory of the world
in which we are lucky enough to be living in one of those regions
which rolled towards the right minimum.  We had a fifty fifty chance.
If we are more greedy, and want to explain the world as we see it as
a dead certainty, then there are several roads that we might follow.
Undoubtedly, there are anthropic arguments which show that human
beings cannot survive in the weak coupling region.
More physically, we can imagine features of the modular potential
which could prevent the universe from reaching weak coupling.
For example, if the potential had the sort of double ridge structure
shown in Fig. 2, with inflation occurring only on the strong coupling ridge,
then the weak coupling side could only be reached by quantum tunneling
after inflation, while the system could roll classically into the true
vacuum.  Since tunneling lifetimes for the moduli are exponentially
longer than the age of the universe \papertwo{}
, the universe would end up in the
true vacuum with overwhelming probability.
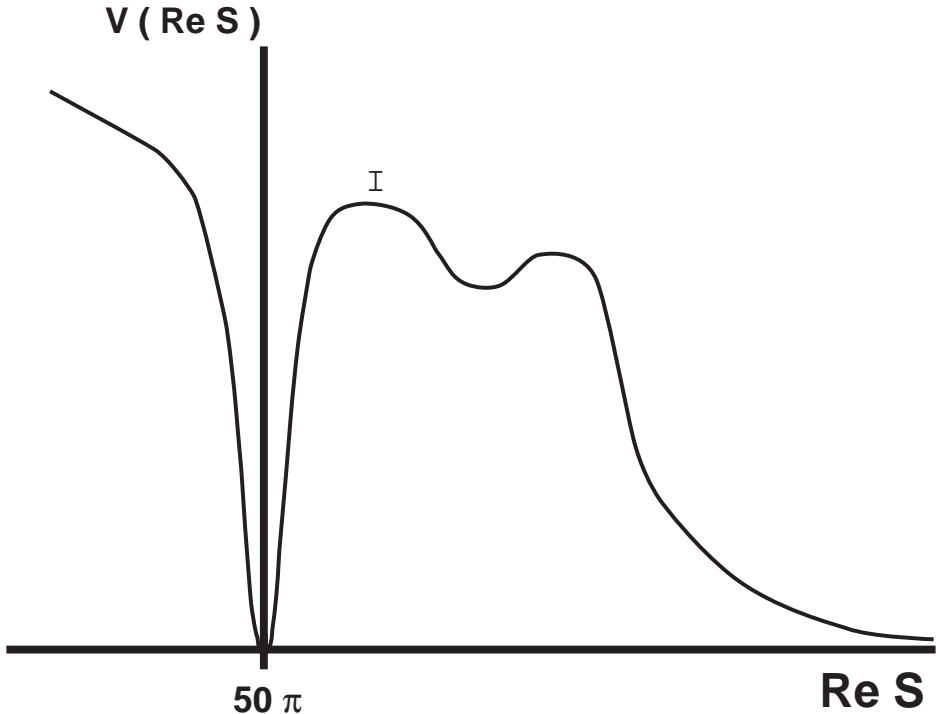
\ifig\ftwo{Inflation off the strong coupling maximum leads either
to classical evolution to the true minimum.  The system can only
get into the weak coupling region with an infinitesimally small
tunneling probability.}
{\epsfysize=8cm \epsfbox{fig1.draw.epsf}}
A more plausible possibility is to assume that the ridge is separated
from the weak coupling region by a region in which the potential is
negative.  The results of \papertwo then imply that after inflation
the system generically tries to roll into the negative potential
region and ends up in a state of contraction instead.  Few if any
trajectories get out to the weak coupling region.  The only generic
metastable behavior of the system after inflation is to come to rest
at the finite coupling vacuum state.

\subsec{Inflationary Fluctuations and Another Model for Superstring
Inflation}

We see then that a few simple assumptions about the nature of the
potential on moduli space lead to a rather robust prediction of an
inflationary
universe which settles into the correct ground state of string theory
after inflation has ended.  We must ask whether this scenario for inflation
passes the crucial tests of any inflationary model, the generation
of density fluctuations of the right magnitude, and proper reheating
of the universe.  Consideration of the first of these tests leads to
an interesting general observation.  We have emphasized that in the
context of natural models, generation of the right fluctuation
amplitude requires that the height of the potential near its
inflationary maximum be about $10^{-10} M_P^4$.
This is quite different from the maximum allowed scale for
SUSY breaking, $10^{-32} M_P^4$.

We can think of two natural explanations for this discrepancy.
The first involves the fact that the potential we are discussing
varies exponentially with the string coupling in the weak coupling region.
Inflation takes place on a ridge in moduli space, while SUSY breaking
is associated with the natural scale of the potential near its minimum.
If, as in Fig.3  the ridge sits at a {\it stronger} value of the
coupling than the minimum, we can expect the potential near the
ridge to be substantially larger than the SUSY breaking scale.
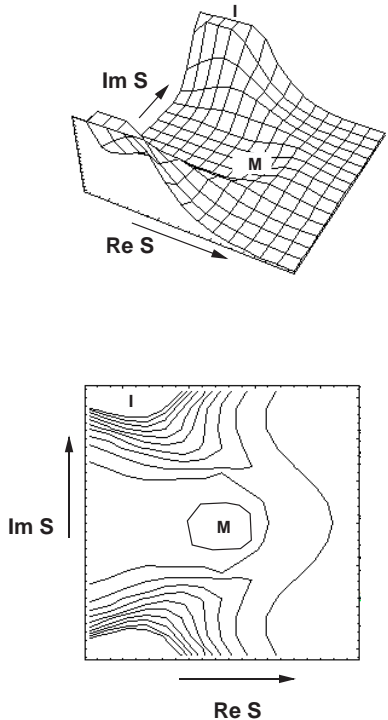
\ifig\fthree{In the one component scenario, the discrepancy between inflation
and SUSY breaking scales is explained because inflation occurs on a ridge I
at stronger coupling than the minimum M.  The potential varies exponentially
with the coupling.}
{\epsfysize=8cm \epsfbox{fig2.draw.epsf}}

Another, more {\it modular}, approach is to assume that inflation
and SUSY breaking are related to two different sectors of the theory.
That is, we assume the existence of two hidden sectors whose
nonperturbative dynamics occurs at the relevant scales.  The first
sector preserves SUSY and has zero vacuum energy.  Its potential
has an inflationary ridge.  The second breaks SUSY, at a lower scale.
This scenario has some attractions.  The ``observed'' value of
the unification scale for the gauge couplings suggests the existence
of a threshold for new physics at about the energy scale needed
to explain the observed microwave fluctuations in terms of inflation.

In \papertwo , we have investigated the scenario in which modular
masses arise from high scale, SUSY preserving dynamics.  There we argue that it
is
unlikely for SUSY preserving dynamics which gives rise to
a vanishing cosmological constant to give mass to the dilaton\foot{Here
we are relying on our religious conviction that
one cannot cancel two different exponentially small effects
in the weak coupling region.}.  If these arguments are
correct, then the model of inflation occurring on the ridge
between strong and weak coupling
will not be compatible with the idea of high scale, SUSY
preserving nonperturbative dynamics.

However, the idea of defect inflation is very robust, and a version of
it also occurs naturally in the {\it two component} scenario.  In this scenario
the superpotential consists of two parts:
\eqn\susysupot{ W = e^{-b S}W_1 (M) + e^{- cS} W_2 (M)}
Here $b < c$, so that, at small values of the coupling and generic
values of $M$, the second term is negligible.
We can think of these two terms as arising from gaugino condensation or similar
nonperturbative dynamics in two disjoint sectors of the hidden gauge group
of superstring theory.
We postulate a minimum of the potential generated by $W_1$, which preserves
both
SUSY and a complex R-symmetry, so that the cosmological constant is zero
\papertwo .
This means that $W_1$ vanishes at the minimum.  As a consequence, the effective
potential for the dilaton would vanish, were it not for the second term in
the superpotential.  When combined with a Kahler potential arising from
nonperturbative
string physics, this term can stabilize the dilaton, but the cosmological
constant will vanish only if SUSY is broken \banksdine.  By assumption, the
minimum
of the first part of the potential is isolated, so that all moduli but the
dilaton
obtain a mass of order $e^{-b S} M_P$.  The SUSY breaking F term has a scale
$e^{-cS} M_P^2$, and the gravitino mass
$e^{-cS}M_P$, which is much smaller than the moduli masses.

In flat space, lagrangians with a superpotential of the form \susysupot
have domain wall solutions.  Consider
any \lq\lq noncontractible loop''
 in moduli space which runs through the minimum
of the potential, $M = 0$.
A field configuration in which $M(z)$ winds around this noncontractible
loop, approaching $M = 0$
as the space coordinate $z$ goes to
 $-\infty$ and $\infty$, would be a metastable
domain wall configuration.
{}.
Our discussion of the preinflationary dynamics
of moduli indicates that many such walls will be produced in the course of
expansion down from energy densities near the Planck scale.
Further, the same sort of mild
fine tuning which produced defect inflation in the scenario
of the previous section will work the same sort of magic here.

The major difference between these two models for explaining the discrepancy
between the inflationary and SUSY scales relates to the problem discovered
in \problems .  The ``fifty percent solution'' advocated above really works
only
if the inflationary saddle lies at an energy density not much higher than
the lowest ridge separating the true minimum from the weak coupling region.
In order to explain the discrepancy between the SUSY and inflationary scales
with a single exponential contribution to the superpotential, we must assume
the energy density at the saddle is $10 ^{22}$ times larger than the energy
density near the SUSY breaking minimum.  Although our system is
multidimensional, one would expect a substantial fraction of this energy to
be diverted into kinetic energy of motion in the direction of weak coupling.
It is likely to
send the system
flying over the miniscule barrier into the weak coupling region\foot{
Actually, the problem described in \problems is not quite as serious
as it was depicted there.  We show in the appendix that for a large class
of rapidly decreasing potentials, the zero mode
of the dilaton coupled
to gravity changes only logarithmically with the energy.  Thus a
decrease of 22 orders of magnitude in energy corresponds to an increase of
$Re{}S$ by about $12$.  This might be small enough to avoid flying
over the barrier.  See the Appendix for further discussion.}
, following the scenario of \problems .

By contrast, the postinflationary fate of the dilaton in the high scale SUSY
preserving scenario, depends on the effective potential for the dilaton
generated by the moduli fields during and just after inflation. The moduli
 are initially
displaced from their minima\foot
{We choose coordinates on moduli space so that the minimum is at $M=0$.}
(because they
begin moving from the top of the domain wall) just after inflation.
If the combined modular-dilaton potential has the form shown in Fig. 1
it will have a local minimum for $S$, $S = S_0 (M)$, for nonzero $M$.
If the $M=0$ end of the \lq\lq groove `` $S_0 (M)$ is on the strong
coupling side of the minimum of the potential generated by $e^{-cS}
W_2 (0)$,
then after inflation, $S$ will evolve toward the vicinity of $S_0 (0)$
and will not have a large velocity along the dilaton direction.
Once the energy density comes down to of order $e^{-2c Re S} M_P^4$, the
second part of the potential will come into play and the dilaton
will evolve towards its minimum.

Thus, the two component approach to modular cosmology seems to be preferred
if we want to both solve the problem of \problems and explain the relative
scales of inflation and SUSY breaking.  As emphasized in \papertwo ,
it also provides us with a very strong hint about where in moduli space
the true vacuum of string theory must lie.  In a way, the results
of the appendix are somewhat disappointing because they mean that this
conclusion is not completely clear cut.  The single exponential
scenario might, depending upon numerical details over which we have
no control at present, also survive these two tests of a successful
superstring cosmology.

\subsec{Reheating - The Postmodern Polonyi Problem}

Reheating has been the bugaboo of supersymmetric inflationary
models almost since the beginning of time.  Typical hidden sector
models contain very light scalar fields with gravitational
strength coupling.  In an inflationary universe model, such fields
inevitably start the post-inflationary period displaced from
their minima by a finite amount.  They remain practically constant
until the Hubble constant falls to a value on the order of their masses.
For natural models, in which the potential takes the form
$M^4 V({\phi\over M_P})$,
this takes place at a time when the total energy density
of the universe is on the order of the energy stored in these
displaced scalars.  From that point on the scalar fields behave like
nonrelativistic matter, and the universe is matter dominated until they
decay.  For weak scale masses and Planck scale couplings, the energy
density at the time of decay is eight orders of magnitude smaller than
that required for nucleosynthesis.

The paper \bkn pointed out the generality of this phenomenon for theories
with hidden sectors coupled to the ordinary world only through
gravitational strength interactions.   There and in \casas{} , it
was also noted that stringy moduli would be prime examples of this
problem.  A number of proposals to solve the PPP have been
made, but it is not clear that any of them are successful.  The problem
of reheating appears to be the most serious obstacle to the construction
of a viable modular cosmology.

\newsec{Conclusions}

In summary, we have generalized the arguments of \frieman , which suggest
that inflation is only compatible with field theoretic naturalness if
there exist scalar fields, like string theory moduli, whose natural range
of variation is the Planck scale.  The COBE and gravitational
radiation constraints on the vacuum energy during inflation put
another obstacle in the way of a successful inflationary scenario.
They imply the existence of an era when the potential energy was
negligible and the total energy density well below the Planck density.
In models of the type we are considering, in which the potential is
bounded, generic initial conditions in this preinflationary era do not set
up the right conditions for inflation, unless domain walls (or possibly
other kinds of topological defects) are formed.

We showed that the dynamics of stringy moduli may give rise to two
different kinds of domain walls, the first associated with the ridge
between strong and weak coupling, and the second with noncontractible
loops in moduli space.  Both can give rise to {\it defect inflation}
if the potential at the top of the wall is flat enough.

Finally, we tried to incorporate into our cosmology an explanation
of the large difference between the inflationary scale and the SUSY breaking
scale.  Of the two models proposed for this purpose, the two component
approach in which the superpotential is the sum of an inflationary, SUSY
preserving piece and an hierarchically smaller piece which leads to
spontaneous SUSY violation solves the dilaton runaway problem posed
in \problems in a more elegant manner.  It also promises a criterion
for determining the correct superstring vacuum state, and gives
a mass much larger than that of the gravitino to all moduli but the dilaton.

Such a modular cosmology would lead to dilaton domination of the universe
from an energy density of about $(10^{11}{} GeV)^4$ down to $(10^{-2}
{} MeV)^4$.  It is not clear how to eliminate the dilatons at an
early enough stage to make the model compatible with classical cosmology.
This seems to be the most serious problem of any modular cosmology.
If it cannot be solved, the present form of string theory will have been
shown to be incompatible with observation.

Cosmological considerations thus lead to an intricate and detailed
set of constraints on the hidden sector of string theory.  Heretofore
this sector has been a black box whose only function was to generate
SUSY breaking.  Modular cosmology has suggested that it should consist of
two sectors with rather distinct properties, which become strong at very
specific energy scales.  Combined with knowledge of the string coupling
at very high energies these scales determine the value of the leading
coefficient in the beta function of each of the two components of the
hidden sector.  We suspect that further study of modular cosmology
will lead to additional constraints on the hidden sector.

Before concluding, we would like to point out that many of the conclusions
of this paper are true quite generally of models of inflation that obey
the field theoretic constraints of naturalness.  Fields whose natural
range of variation is the Planck mass, and whose couplings to ordinary
matter are suppressed by the Planck scale,
seem to be required for inflation.  The scale of the potential
during inflation is fixed to be about $10^{-10} M_P^4$ in such theories,
by the COBE data on microwave background fluctuations.  In any case,
constraints on primordial gravitational waves tells us that the vacuum
energy density during inflation cannot be larger than this.
As a consequence, there will be a \lq\lq window'' of cosmic history
in which we might expect a semiclassical treatment of the universe
to be valid, but during which the potential of the inflaton fields
is negligible.  We saw in models of the type studied in this paper,
such a pre-inflation era does not
lead to the natural initial conditions for inflation, unless the
model supports \lq\lq flat topped defects'' (FTD's) .
Thus, the necessity for and the robustness of defect inflation
may be general features of a large class of inflationary models,
not just those motivated by superstrings.  We note however that in
models where inflation occurs at large field strengths, or from a
metastable minimum of the effective potential, such defects may not be
necessary.
\vfill\eject
\centerline{\bf Acknowledgements}

GM would like to thank the physics department of Rutgers University for
its hospitality and J Horne for interesting discussions about this work.
This work originated while PJS was a Visiting Professor
at the Theory Group at Rutgers University
and was completed while PJS as a Visiting Fellow at
IAS in Princeton and ITP in Santa Barbara.
PJS would like to thank Rutgers, IAS and ITP for
their support and T. Banks, S. Shenker, N. Seiberg
and D. Friedan for their generous hospitality during
his visit to Rutgers.
TB would like to thank ITP Santa Barbara, the Aspen Center for Physics,
and the Particle Physics Department of the Weizmann Institute of Science
for their hospitality while parts of this work were carried out.  He
would also like to thank L.Randall, S.Thomas, E.Witten and
K.Intriligator for discussions.
MB would like to thank the Weizmann Institute of Science for its
hospitality.
SS would like to thank P. Binetruy and E. Witten for helpful
discussions.

\vfill\eject
\centerline{\bf Appendix}

In dimensionless units, the equations of motion for a single scalar field
$S$ (the inverse string coupling),
 coupled to gravity in a homogeneous isotropic, spatially
flat universe
are
\eqn\eoma{\ddot{S} + 3 H\dot {S} + V^{\prime} (S) = 0}
\eqn\eomb{H = \sqrt{E} = \sqrt{\ha (\dot{S} )^2 + V(S)}}
They imply
\eqn\eomc{\dot{E} = - 6\sqrt{E}(E - V)}
and
\eqn\eomd{\dot{S} =\sqrt{2(E - V)}}

We have chosen a canonical kinetic term for $S$.  The purpose of the
present appendix is to explore the evolution of the coupling starting
from an inflationary ridge at very high energy density in the one
component scenario for modular cosmology.  In this scenario, the
difference between the inflationary energy density and the SUSY scale is
attributed to an exponential of the inverse coupling in the expression
for the potential.  Inflation occurs in a more strongly coupled region
of field space than SUSY breaking. We are trying to determine
whether the dilaton field, whose initial energy is very high,
is likely to simply sail over the small barrier separating
the finite coupling minimum from the the weak coupling region.  We are
thus interested in a region of field space in which, according to the
philosophy of \banksdine , the Kahler potential for $S$ is unknown.
It would be wrong to use the weak coupling expression for the Kahler
potential in this region.  Instead, we expand the Kahler potential
around the initial point, assuming, consistent with \banksdine that
it is not a rapidly varying function in this region.  Only the
superpotential is rapidly varying.  What we will show below is that,
despite, (indeed, because of) the steepness of the potential, the
dilaton moves a distance which is only logarithmic in the decrease in
energy density.  Thus, keeping only the first term in the expansion of
the Kahler potential around the initial point is a reasonable
approximation.

In an expanding universe, $E$ is decreasing and will eventually reach zero.
It is convenient to define $E \equiv e^{-u}$.  $u$ will go to $+\infty$
asymptotically.  Divide equation \eomd by \eomc and rewrite everything
in terms of $u$.
The result is
\eqn\eome{S_u = {1\over 3\sqrt{2(1 - e^u V)}}}.
$e^u V$ must of course remain bounded by $1$.
In general, if $V$ is rapidly
decreasing with $S$, $e^u V$ will in
fact go to zero rapidly as $S$ increases.
Thus, the logarithmic derivative of $S$ becomes constant, so that the
dilaton increases only logarithmically as $E$ decreases.

It is easy to see that power law increase of $S$ as a function of $E$
is only consistent if the potential has a decreasing power law form.
In fact, the boundary between the behavior we described above and systems
for which the potential become important asymptotically, is the
exponentially decreasing potential $e^{-b S}$.  The expected form
is precisely exponential.  Note that if we had used the weak coupling
form of the Kahler potential, the canonically normalized field would be
the logarithm of $S$ and the potential would have been even steeper than
this.

The ratio between inflationary and SUSY energy densities is about
$10^{22}$, with natural log about $51$.  The nominal distance between
the inflationary plateau and the SUSY breaking vacuum state for $S$ may
be crudely calculated as follows:  The scale of SUSY breaking
$\sim 10^{-8} M_P$ should be explained as $e^{-bS_{vac}} M_P$, where
$S_{vac}$
is determined by the \lq\lq observed'' value of the unified coupling to
be about $50\pi$.  The position of the inflationary plateau is
approximately determined by $e^{-b S_I} M_P \sim 10^{-2.5} M_P$.
Thus $S_I \sim {2.5\over 8}S_{vac}$ and $S_{vac} - S_I \sim 35\pi$ .
So the question of whether or not the dilaton flies out to weak coupling
is whether $51 C > 35\pi$, where $C$ is the coefficient of logarithmic
increase in the above solution,$S \rightarrow - C ln E$.  Clearly the
answer depends on $C$.  Even more clearly, our imperfect knowledge of
the potential and kahler potential on moduli space precludes our giving
a convincing answer to this question at present.

We conclude that the dilaton runaway problem discovered in \problems ,
which is not resolved in our one component scenario, may not really be a
problem. However the resolution of this question depends on numerical
details of the effective lagrangian for moduli over which we have no
control at present.  By contrast, the successful resolution of dilaton
runaway in the two component scenario depends only on the qualitative
assumption that the \lq\lq dilaton groove'' in the inflationary
potential for moduli terminates at a value of the coupling stronger than
the position of the barrier between strong and weak coupling in the SUSY
breaking potential.

\listrefs
\end